\documentstyle[aps,epsf,preprint]{revtex}

\begin{document}

\title{
Quasiparticle RPA
with finite rank approximation for Skyrme
interactions}
\author{A.P. Severyukhin}
\address{Bogoliubov Laboratory of Theoretical Physics, Joint Institute for
Nuclear Research, 141980 Dubna, Moscow region, Russia}
\author{Ch. Stoyanov}
\address{Institute for Nuclear Research and Nuclear Energy, boul.
Tzarigradsko Chaussee 72, 1784 Sofia, Bulgaria}
\author{V.V. Voronov}
\address{Bogoliubov Laboratory of Theoretical Physics, Joint Institute for
Nuclear Research, 141980 Dubna, Moscow region, Russia}
\author{Nguyen Van Giai}
\address{Institut de Physique
Nucl\'eaire, Universit\'e Paris-Sud, F-91406 Orsay Cedex, France}

\maketitle
\begin{abstract}
A finite rank separable approximation for the particle-hole RPA
calculations with Skyrme interactions is extended to take into
account the pairing. As an illustration of the method energies and
transition probabilities for the quadrupole and octupole
excitations in some O, Ar, Sn and Pb isotopes are calculated.  The
values obtained within our approach are very close to those that
were calculated within QRPA with the full Skyrme interaction. They
are in reasonable agreement with experimental data.
\end{abstract}

\section{Introduction}
Many properties of the nuclear states can be described
within the random phase approximation (RPA)\cite{R70,BM75,Schuck,solo}.
Among many microscopic nuclear models aiming at a
description of the properties of nuclear excitations
the most consistent model employs an effective interaction which
can describe, throughout the periodic table, the ground states in the
framework of the Hartree-Fock (HF) approximation and the excited states in
 the small amplitude limit of the time-dependent HF, or the random phase approximation (RPA).
The Gogny's interaction\cite{gogny} and the
Skyrme-type interactions\cite{vau72} are very popular now.
Such models are quite successful not
only for predicting accurately nuclear ground state
properties\cite{floc78,doba96} but also for calculating the main features of
giant resonances in closed-shell nuclei\cite{colo,colo1} and
single-particle strengths near
closed shells\cite{bern}. Taking into account the pairing effects
enables one to reproduce also many properties of collective states in
open-shell nuclei \cite{KG00,Khan00,colo2,colo3}.

It is well known that due to the anharmonicity of vibrations there is a
coupling between one-phonon and more complex states \cite{BM75,solo}.
The main difficulty is that the
complexity of calculations beyond standard RPA (e.g., for
studying damping mechanisms of collective excitations) increases rapidly
with the size of the configuration space and one has to work within limited
spaces.
From another point of view
more phenomenological models that assume some simple
separable form for the residual nucleon-nucleon interaction while the mean
field is modelized by an empirical potential well allow
one to calculate nuclear excitations in very large configuration
spaces since there is no need to diagonalize matrices whose dimensions grow
with the size of configuration space.
The well-known quasiparticle-phonon model (QPM) of Soloviev
et al.\cite{solo} belongs to such a model.
Very detailed predictions can be made by QPM
for nuclei away from closed shells\cite{gsv}.

The possibility to solve easily the RPA problem in a large
configuration space when the residual particle-hole (p-h)
interaction is separable was the motivation for proposing in our
previous work \cite {gsv98} a finite rank approximation for the
p-h interaction resulting from Skyrme-type forces.
Thus, the self-consistent mean field can be calculated
in the standard way with the original Skyrme interaction whereas the RPA
solutions would be obtained with the finite rank approximation to the p-h
matrix elements.
It was found that the finite rank approximation reproduces reasonably well
the dipole and quadrupole strength distributions in Ar isotopes.

In the present work, we extend the finite rank RPA calculations
to take into account pairing effects.
As an application we present results of
calculations for low-lying $2^+$ and $3^-$ states in some O, Ar, Sn and
Pb isotopes. This paper is organized as
follows: in Section II we sketch our method for constructing a finite rank
interaction for the quasiparticle RPA (QRPA) case.
In Section III we discuss details of calculations and show how
this approach can be applied to treat different multipole states
in wide excitation  energy regions.
Results of calculations
for characteristics of the quadrupole and octupole states in some nuclei
are given in Section IV. Conclusions are drawn in Section~V.

\section{Hamiltonian of the model and QRPA}

We start from the effective Skyrme interaction\cite{vau72}
and  use the notation of Ref.\cite{sg81} containing explicit
density dependence and all spin-exchange terms rather than the original
form of Ref\cite{vau72} where density dependence at the HF level was
introduced by a three-body contact force and where some spin-exchange terms
were dropped. The exact p-h residual interaction $\tilde V_{res}$
corresponding to the Skyrme force and including both direct and exchange
terms can be obtained as the second derivative of the energy density
functional with respect to the density\cite{ber75}.
Following our previous paper\cite{gsv98} we  simplify $\tilde V_{res}$ by
approximating it by its Landau-Migdal form in the momentum space:

\begin{eqnarray}
V_{res}({\bf k}_1,{\bf k}_2)=N_0^{-1}\sum_{l=0}^1\left[ F_l+G_l%
{\bf \sigma}_1 \cdot{\bf \sigma}_2+(F_l^{^{\prime }}+G_l^{^{\prime
}}{\bf \sigma }_1 \cdot{\bf \sigma}_2){\bf \tau}_1 \cdot{\bf \tau%
}_2\right] P_l(\frac{{\bf  k}_1 \cdot{\bf k}_2}{k_F^2}),
\label{eq1}
\end{eqnarray}

where ${\bf k}_i$, ${\bf \sigma}_i$ and ${\bf \tau}_i$ are the
nucleon momentum, spin and isospin operators,
and \\ $N_0 = 2k_Fm^{*}/\pi^2\hbar^2$ with $k_F$ and $m^{*}$ standing for the
Fermi momentum and nucleon effective mass.
For Skyrme interactions all Landau parameters with $l > 1$ are zero. Here,
we keep only the $l=0$ terms in $V_{res}$ and in the coordinate
representation one can write it in the following form:

\begin{eqnarray}
V_{res}({\bf r}_1,{\bf r}_2)=N_0^{-1}\left[ F_0(r_1)+G_0(r_1)
{\bf \sigma}_1 \cdot{\bf \sigma}_2+(F_0^{^{\prime
}}(r_1)+G_0^{^{\prime }}(r_1){\bf \sigma }_1 \cdot{\bf \sigma}_2){\bf
\tau }_1 \cdot{\bf \tau }_2\right] \delta ({\bf r}_1-{\bf r
}_2)  \label{eq2}
\end{eqnarray}

The expressions for
$F_0, G_0, F^{'}_0, G^{'}_0$ in terms of the Skyrme force parameters can
be found in Ref.\cite{sg81}. Because of
the density dependence of the interaction the Landau parameters of
Eq.(\ref{eq2}) are functions of the coordinate ${\bf r}$.
In what follows we use the second quantized representation
and $V_{res}$ can be written as:
\begin{eqnarray}
\hat V_{res} & = & \frac 12\sum_{1234}V_{1234}:a_1^{+}a_2^{+}a_4 a_3:
\end{eqnarray}
where $a^+_1$ ($a_1$) is the particle creation (annihilation) operator
and $1$ denotes the quantum numbers $(n_1l_1j_1m_1)$,

\begin{eqnarray}
V_{1234} = \int \phi^*_1({\bf r}_1)\phi^*_2({\bf r}_2)
V_{res}({\bf r}_1,{\bf r}_2)\phi_3({\bf r}_1)
\phi_4({\bf r}_2) {\bf dr}_1{\bf dr}_2 ,
\end{eqnarray}

\begin{eqnarray}
V_{1234}=\sum_{JM}\hat J^{-2}
(-)^{K}\langle j_1m_1j_3-m_3 \mid J-M\rangle
\langle j_2m_2j_4-m_4\mid JM\rangle V_{1234}^J,
\end{eqnarray}
where $K=J+j_3+j_4-M-m_3-m_4$ and
\begin{eqnarray}
V_{1234}^J &=&\langle j_1||Y_J||j_3\rangle \langle
j_2||Y_J||j_4\rangle I_M(j_1j_2j_3j_4)- \nonumber\\
&&\sum_{L=J,J\pm 1}\langle j_1||T_{JL}||j_3\rangle \langle
j_2||T_{JL}||j_4\rangle I_S(j_1j_2j_3j_4).
\end{eqnarray}
In the above equation, $\langle j_1 \vert\vert Y_{J} \vert \vert j_3 \rangle$ is
the reduced matrix element of the spherical harmonics $Y_{J \mu}$,
$\hat J = \sqrt {2J+1}$,
$T_{J L}^{M}(\hat r, \sigma) = [Y_L \times \sigma]_{J}^{M}$
and $I_M(j_1j_2j_3j_4)$, $I_{S}(j_1j_2j_3j_4)$ are the radial integrals:

\begin{eqnarray}
I_M(j_1j_2j_3j_4)=N_0^{-1}\int_0^\infty \left( F_0(r)+F_0^{\prime }(r){\bf
\tau }_1 \cdot{\bf \tau }_2\right)
u_{j_1}(r)u_{j_2}(r)u_{j_3}(r)u_{j_4}(r)\frac{dr}{r^2}, 
\end{eqnarray}

\begin{eqnarray}
I_S(j_1j_2j_3j_4)=N_0^{-1}\int_0^\infty \left( G_0(r)+G_0^{\prime }(r){\bf
\tau }_1 \cdot{\bf \tau }_2\right)
u_{j_1}(r)u_{j_2}(r)u_{j_3}(r)u_{j_4}(r)\frac{dr}{r^2},  
\end{eqnarray}
where the radial wave functions $u(r)$ are related to the
HF single-particle wave functions:
\begin{eqnarray}
\phi_{i,m}(1) & = & \frac {u_{i}(r_1)}{r_1} {\cal Y}_{l_i,j_i}^{m}
(\hat {r_1},\sigma_1)~.
\end{eqnarray}

As it is shown in \cite{gsv98} the radial integrals can be calculated
accurately by choosing a large enough cutoff radius $R$
and using a $N$-point integration Gauss formula with abscissas and weights
${r_k,w_k}$. Thus, the residual interaction can be presented as a sum of
$N$ separable terms.

So we employ the hamiltonian
including an average nuclear HF field,
pairing interactions, the isoscalar and
isovector particle--hole (p--h)
residual forces in a finite rank $N$:

\begin{eqnarray}
H &=&\sum\limits_\tau \left( \left. \sum\limits_{jm}\right. ^\tau
(E_j-\lambda _\tau )a_{jm}^{\dagger }a_{jm}-\frac 14V_\tau
^{(0)}:P_0^{\dagger }\,(\tau )P_0\,(\tau ):-\right.\nonumber \\
&&\frac 12\sum_{k=1}^{N}\sum_{q=\pm 1}\sum\limits_{\lambda \mu
}\left[ \left( \kappa _0^{(M,k)}+q\kappa _1^{(M,k)}\right) :M_{\lambda \mu
}^{\left( k\right) +}(\tau )M_{\lambda \mu }^{\left( k\right) }(q\tau
):+\right.  \nonumber \\
&&\left. \left. \sum_{L=\lambda ,\lambda \pm 1}\left( \kappa
_0^{(S,k)}+q\kappa _1^{(S,k)}\right) :S_{\lambda L\mu }^{\left( k\right)
+}(\tau )S_{\lambda L\mu }^{\left( k\right) }(q\tau ):\right] \right)
\label{eq9},
\end{eqnarray}
We sum over the proton($p$) and neutron($n$) indexes and the notation $
\{\tau =(n,p)\}$ is used. A change $\tau \leftrightarrow -\tau $ means a
change $p\leftrightarrow n$.
The single-particle states are specified by the
quantum numbers $(jm)$, $E_j$ are the single-particle energies,
$\lambda_\tau $ the chemical potentials. $V_\tau ^{(0)}$  is the
interaction
strength in the particle-particle channel, $\kappa ^{(M k)}$ ($\kappa ^{(S k)}$)
are the multipole (spin-multipole) interaction strengths in the p--h channel and
they can be expressed via the Landau parameters as:

\begin{equation}
\left(
\begin{array}{c}
\kappa _0^{(M,k)} \\
\kappa _1^{(M,k)} \\
\kappa _0^{(S,k)} \\
\kappa _1^{(S,k)}
\end{array}
\right) =-N_0^{-1}\frac{Rw_k}{2r_k^2}\left(
\begin{array}{c}
F_0(r_k) \\
F_0^{\prime }(r_k) \\
G_0(r_k) \\
G_0^{\prime }(r_k)
\end{array}
\right)  
\end{equation}

The
monopole pair creation, the multipole and spin-multipole operators
entering the normal products in Eq.(\ref{eq9}) are defined as follows:

\begin{equation}
P_0^{+}\,(\tau )=\,\left. \sum_{jm}\right. ^\tau
(-1)^{j-m}a_{jm}^{+}a_{j-m}^{+},
\end{equation}

\begin{equation}
M_{\lambda \mu }^{\left( k\right) +}\left( \tau \right) \,=\,\hat \lambda
^{-1}\left. \sum_{jj^{^{\prime }}mm^{^{\prime }}}\right. ^\tau
(-1)^{j+m}\langle jmj^{^{\prime }}-m^{^{\prime }}\mid \lambda \mu \rangle
f_{j^{\prime }j}^{(\lambda k)}(\tau )a_{jm}^{+}a_{j^{^{\prime }}m^{^{\prime
}}},
\end{equation}

\begin{equation}
S_{\lambda L\mu }^{\left( k\right) +}\left( \tau \right) \,=\,\hat \lambda
^{-1}\,\left. \sum_{jj^{^{\prime }}mm^{^{\prime }}}\right. ^\tau
(-1)^{j+m}\langle jmj^{^{\prime }}-m^{^{\prime }}\mid \lambda \mu \rangle
f_{j^{\prime }j}^{(\lambda Lk)}(\tau )a_{jm}^{+}a_{j^{^{\prime }}m^{^{\prime
}}},
\end{equation}

where $f_{j^{^{\prime }}j}$ are the single particle radial
matrix elements of the multipole and spin-multipole operators:

\begin{equation}
f_{j_1j_2}^{(\lambda k)}=u_{j_1}(r_k)u_{j_2}(r_k)i^\lambda \langle
j_1||Y_\lambda ||j_2\rangle
\end{equation}

\begin{equation}
f_{j_1j_2}^{(\lambda Lk)}=u_{j_1}(r_k)u_{j_2}(r_k)i^L\langle j_1||T_{\lambda
L}||j_2\rangle
\end{equation}
One can see that the hamiltonian (\ref{eq9}) has the same form as the QPM
hamiltonian with $N $ separable terms, but in contrast to the QPM all
parameters of this hamiltonian are expressed through parameters of the
Skyrme forces.

In what follows we work in the quasiparticle  representation defined by
the canonical Bogoliubov transformation:
\begin{equation}
a_{jm}^{+}\,=\,u_j\alpha _{jm}^{+}\,+\,(-1)^{j-m}v_j\alpha _{j-m}.
\label{B}
\end{equation}
The hamiltonian (\ref{eq9}) can be represented in terms of bifermion
quasiparticle operators and their conjugates \cite{solo}:
\begin{equation}
B(jj^{^{\prime }};\lambda \mu )\,=\,\sum_{mm^{^{\prime }}}(-1)^{j^{^{\prime
}}+m{^{\prime }}}\langle jmj^{^{\prime }}m^{^{\prime }}\mid \lambda \mu
\rangle \alpha _{jm}^{+}\alpha _{j^{^{\prime }}-m^{^{\prime }}},
\end{equation}
\begin{equation}
A^{+}(jj^{^{\prime }};\lambda \mu )\,=\,\sum_{mm^{^{\prime }}}\langle
jmj^{^{\prime }}m^{^{\prime }}\mid \lambda \mu \rangle \alpha
_{jm}^{+}\alpha _{j^{^{\prime }}m^{^{\prime }}}^{+}.
\end{equation}
We introduce the phonon creation operators
\begin{equation}
Q_{\lambda \mu i}^{+}\,=\,\frac 12\sum_{jj^{^{\prime }}}\left( X
_{jj^{^{\prime }}}^{\lambda i}\,A^{+}(jj^{^{\prime }};\lambda \mu
)-(-1)^{\lambda -\mu }Y _{jj^{^{\prime }}}^{\lambda i}\,A(jj^{^{\prime
}};\lambda -\mu )\right).
\end{equation}
where the index $\lambda $ denotes total angular momentum and $\mu $ is
its z-projection in the laboratory system.
One assumes that the ground state  is the QRPA phonon vacuum
$\mid 0\rangle $,\\ i.e. $Q_{\lambda \mu i}\mid 0\rangle\,=0$.
We define the excited states for this approximation by
$Q_{\lambda\mu i}^{+}\mid0\rangle$.
For the QRPA the following relation is valid:
\begin{equation}
\langle 0\mid \,[Q_{\lambda \mu ,i},Q_{\lambda ^{^{\prime }}\mu ^{^{\prime
}},i^{^{\prime }}}^{+}]\,\mid 0\rangle \,=
\delta_{\lambda \lambda^{\prime}}\delta_{\mu\mu^{\prime}}\frac 12\sum_{jj^{^{\prime }}}\left( X
_{jj^{^{\prime }}}^{\lambda i}\,X _{jj^{^{\prime }}}^{\lambda i^{^{\prime
}}}\,-\, Y _{jj^{^{\prime }}}^{\lambda i} Y _{jj^{^{\prime
}}}^{\lambda i^{^{\prime }}}\right)
\end{equation}

The quasiparticle energies ($\varepsilon _j$),
the chemical potentials ($\lambda_\tau $), the energy gap and the coefficients
$u$,$v$ of the  Bogoliubov transformations
(\ref{B}) are determined
from  the BCS equations with the single-particle spectrum that is calculated
within the HF method with the effective Skyrme interaction. Making use
of the linearized equation-of-motion approach \cite{R70}:

\begin{equation}
\langle 0|\left[ \delta Q_{\lambda \mu i},\left[ H,Q_{\lambda \mu
i}^{+}\right] \right] \mid 0\rangle =\omega _{_{\lambda i}}\langle 0|\left[
\delta Q_{\lambda \mu i},Q_{\lambda \mu i}^{+}\right] \mid 0\rangle,
\end{equation}
with the condition:
\begin{equation}
\langle 0\mid \,[Q_{\lambda \mu i},Q_{\lambda \mu i^{^{\prime }}}^{+}]\,\mid
0\rangle \,=\delta _{ii^{^{\prime }}},
\label{eq12}
\end{equation}

one can derive the QRPA equations \cite{Schuck,solo}:
\begin{equation}
\label{eq14}
\left(
\begin{tabular}{ll}
${\cal A}$ & ${\cal B}$ \\
${- \cal B}$ & ${- \cal A}$%
\end{tabular}
\right) \left(
\begin{tabular}{l}
$ X $ \\
$ Y $%
\end{tabular}
\right) =w \left(
\begin{tabular}{l}
$ X $ \\
$ Y $%
\end{tabular}
\right).
\end{equation}

In QRPA problems there appear two types of interaction matrix elements,
the $A^{(\lambda)}_{(j_1j_1^{\prime})_{\tau}(j_2j_2^{\prime})_{q\tau}}$
matrix related to forward-going graphs and the
$B^{(\lambda)}_{(j_1j_1^{\prime})_{\tau}(j_2j_2^{\prime})_{q\tau}}$
matrix related to backward-going graphs. For our case we get the following
expressions:

\begin{eqnarray}
A_{(j_1\geq j_1^{\prime })_\tau (j_2\geq j_2^{\prime })_{q\tau }}^{(\lambda
)} &=&\varepsilon _{j_1j_1^{\prime }}\delta _{j_2j_1}\delta _{j_2^{\prime
}j_1^{\prime }}\delta_{q1}-\hat \lambda ^{-2}\left( 1+\delta
_{j_2j_2^{\prime }}\right) ^{-1}\times  \nonumber \\
&&\sum_{k=1}^{N}\left[ (\kappa _0^{(M,k)}+q\kappa
_1^{(M,k)})u_{j_1j_1^{\prime }}^{(+)}f_{j_1j_1^{\prime }}^{(\lambda k)}(\tau
)u_{j_2j_2^{\prime }}^{(+)}f_{j_2j_2^{\prime }}^{(\lambda k)}(q\tau )+\right.
\nonumber \\
&&\left. (\kappa _0^{(S,k)}+q\kappa _1^{(S,k)})u_{j_1j_1^{\prime
}}^{(-)}f_{j_1j_1^{\prime }}^{(\lambda \lambda k)}(\tau )u_{j_2j_2^{\prime
}}^{(-)}f_{j_2j_2^{\prime }}^{(\lambda \lambda k)}(q\tau )\right],
\end{eqnarray}

\begin{eqnarray}
B_{(j_1\geq j_1^{\prime })_\tau (j_2\geq j_2^{\prime })_{q\tau }}^{(\lambda
)} &=&-\hat \lambda ^{-2}\left( 1+\delta _{j_2j_2^{\prime }}\right)
^{-1}\times  \nonumber \\
&&\sum_{k=1}^{N}\left[ (\kappa _0^{(M,k)}+q\kappa
_1^{(M,k)})u_{j_1j_1^{\prime }}^{(+)}f_{j_1j_1^{\prime }}^{(\lambda k)}(\tau
)u_{j_2j_2^{\prime }}^{(+)}f_{j_2j_2^{\prime }}^{(\lambda k)}(q\tau )-\right.
\nonumber \\
&&\left. (\kappa _0^{(S,k)}+q\kappa _1^{(S,k)})u_{j_1j_1^{\prime
}}^{(-)}f_{j_1j_1^{\prime }}^{(\lambda \lambda k)}(\tau )u_{j_2j_2^{\prime
}}^{(-)}f_{j_2j_2^{\prime }}^{(\lambda \lambda k)}(q\tau )\right],
\end{eqnarray}
 where $
\varepsilon _{jj^{\prime }}=\varepsilon _j+\varepsilon _{j^{\prime }}
$
and
$
u_{jj^{^{\prime }}}^{(\pm )}=u_jv_{j^{^{\prime }}}\pm \,v_ju_{j^{^{\prime
}}}
$.

One can find a prescription how to solve this system and to find
the eigen-energies and phonon amplitudes in  Appendix A (see also \cite{gsv98}).
The matrix dimensions never exceed $4N \times 4N$ independently
of the configuration space size.
The derived equations
have the same form as the QRPA equations in the QPM \cite{solo,sol89},
but the
single-particle spectrum and parameters of the p-h residual interaction are
calculated making use of the Skyrme forces.

\section {Details of calculations}

In this work we use generally
the standard parametrization SIII \cite{be75} of the Skyrme force.
Some examples of calculations with other parameter sets are
presented in the next section. Spherical symmetry is assumed for
the HF ground states. The pairing constants $V^0_{\tau}$ are fixed to
reproduce the odd-even mass difference of neighboring nuclei. As a
result constant pairing gaps have values that are very close to
$\Delta=12.0 A^{-1/2}$ besides a case of semimagic nuclei. It is well
known \cite{KG00,colo2} that the constant gap approximation leads to
an overestimating of occupation probabilities for subshells that are far
from the Fermi level and it is necessary to introduce a cut-off in the
single-particle space. Above this cut-off subshells don't participate in
the pairing effect. In our calculations we choose the BCS subspace
to include all subshells lying below 5 MeV.
In order to perform RPA calculations, the single-particle continuum is
discretized \cite{BG77} by diagonalizing the HF hamiltonian on a basis
of twelve
harmonic oscillator shells and cutting off the single-particle spectra at the
energy of 190 MeV. This is sufficient to exhaust practically all the
energy-weighted sum rule.
As it was shown in our previous calculations \cite{gsv98} we have
adopted the value $N$=24 for the finite rank approximation for the
dipole
and quadrupole excitations in Ar isotopes. Increasing the mass
number and the multipolarity of excitations
demands an increase of the rank to keep the calculations accurate.
Our investigations enable us to conclude that $N$=45 is enough for
multipolarities $\lambda \le 3 $ in nuclei with $A\le 208$.
Increasing $N$, for example, up to $N$=60 in $^{208}$Pb changes results
for energies
and transition probabilities not more than by 1\%, so all calculations
in what follows have been done with  $N$=45.  Our calculations show
that, for the normal parity states one can neglect the spin-multipole
interactions  as a rule and this reduces by a factor 2
the total matrix dimension. For example, for the octupole
excitations in $^{206}$Pb we need to invert a matrix having
a dimension 2N=90 instead of diagonalizing a $1376\times1376$ matrix as
it would be without the finite rank approximation.
For light nuclei the reduction of matrix dimensions
due to the finite rank approximation is 3 or 4.
So, for heavy nuclei our approach gives a large gain in comparison
with an exact diagonalization.

The Landau parameters $F_0$, $G_0$, $F_0^{'}$, $G_0^{'}$ expressed in terms of
the Skyrme force parameters \cite{sg81} depend on $k_F$.
As it is pointed out in our previous work \cite{gsv98} one needs to
adopt some effective value for $k_F$ to give an accurate representation
of the original p-h Skyrme interaction.
To fix the effective values of $k_F$ for
the Landau parameters we use
the self-consistency relation \cite{Mig}.
From the following set of equations:
\begin{equation}
\frac 12\sum_\tau \frac{\delta U_\tau ^{self}}{\delta \rho _\tau }%
=N_0^{-1}\left( F_0+F_0^{^{\prime }}\right)
\end{equation}
\begin{equation}
\frac 12\sum_\tau \frac{\delta U_\tau ^{self}}{\delta \rho _{-\tau }}%
=N_0^{-1}\left( F_0-F_0^{^{\prime }}\right)
\end{equation}
one can get the average field potential corresponding to such a choice
of the residual interaction:
\begin{eqnarray}
U_\tau ^{self}  &=&\rho \left( t_0\left( 1+\frac{x_0}2\right)
+\frac{k_F^2}4\left( t_1\left( 2+x_1\right) +t_2\left( 2+x_2\right) \right)
\right) - \\
&&\rho _\tau \left( t_0\left( \frac 12+x_0\right) -\frac{k_F^2}4\left(
t_2\left( 1+2x_2\right) -t_1\left( 1+2x_1\right) \right) \right) +  \nonumber
\\
&&\frac 1{24}t_3\left( \rho ^{\alpha +1}\left( 2+\alpha \right) \left(
2+x_3\right) -\left( 1+2x_3\right) \times \right.   \nonumber \\
&&\left. \left( 2\rho ^\alpha \rho _\tau +\alpha \rho ^{\alpha -1}\left(
\rho _n^2+\rho _p^2\right) \right) \right)   \nonumber.
\end{eqnarray}

This potential can be compared to $(m^*/m) U^{HF}(r)$ which is the
leader term of a local equivalent potential in Skyrme-HF approach.
It is possible to evaluate the effective value $k_F$ for every nucleus.
One can show that this value is larger than the nuclear matter value in
order to compensate for the effects of the neglected terms $F_1$ and
$G_1$.
To calculate the dipole strength distributions
we choose $k_F$ so that the spurious
isoscalar dipole mode appears at zero excitation energy. The strongest
renormalization of the $k_F$ values in comparison with the nuclear
matter value takes place in light nuclei. For $^{208}$Pb the effective
value $k_F$ becomes rather close to the nuclear matter one.

\section{Results of calculations}

As a first example we examine the multipole strength distributions
in $^{36}$Ar.
The calculated strength distributions are displayed in Fig.1.
For the giant dipole resonance (GDR) and giant quadrupole resonance
(GQR) QRPA gives results that are very similar to our previous calculations
with the particle-hole RPA \cite{gsv98} because the influence of
pairing on the giant resonance properties is weak. This is not the case
for the first $2^+$ and $3^-$ states that will be discussed later. For
the GDR energy centroid we get $E_c=19.9$ MeV and this value is rather
close to the empirical systematics \cite{VW87} $E_c=(31.2 A^{-1/3} +
20.6 A^{-1/6})$ MeV. The isoscalar GQR energy centroid is equal to
$E_c=18.8$ MeV that can be compared with the empirical value
$E_c=63 A^{-1/3}=19.1$ MeV. For the isovector GQR our calculation gives
$E_c=30.5$ MeV that is about 10\% less than predicted by the empirical
systematics. It is worth to mention that experimental data for the giant
resonances in light nuclei are very scarce.

Results of our calculations and experimental data \cite{Ram01} for the
$2^+_1$ state energies and transition probabilities $B(E2)$ in several
nuclei are shown in Table 1.
One can see that there is a satisfactory agreement with
experimental data. Results of our calculations for O and Ar isotopes
are close to those of QRPA with Skyrme forces \cite{KG00,Khan01}
and all calculations fail to reproduce the B(E2) value in
$^{18}$O. Making use of the SGII interaction\cite{sg81} improves the
description for the O isotopes and gives practically the same results
for the Ar isotopes, but for Sn and Pb isotopes the results become
much worse. Calculations with the SkI4 force \cite{SkI4} don't change the
above conclusions.
The evolution of the B(E2)-values in the Ar isotopes
demonstrates clearly the pairing effects. The experimental and
calculated B(E2)-values in $^{38}$Ar are three times less than those in
$^{36,40}$Ar. The neutron shell closure leads to the vanishing of
the neutron pairing  and a reduction of the proton gap. As a result
there is a remarkable reduction of the E2 transition probability in
$^{38}$Ar. Some overestimate of the energies indicates
that there is room for two-phonon effects.
Indeed, it was found in calculations performed within the QPM
for $^{208}$Pb \cite{SSV83} that the two-phonon configurations
can shift down the $2^+_1$ energy by more than 1 MeV. The
B(E2)-value reduction is about 10\% in this case. The study of the
influence of two-phonon configurations on properties of the low-lying
states within our approach is in progress now.

Results of our calculations for the $3^-_1$ energies and the transition
probabilities B(E3) are compared with experimental data \cite{Sp02} in
Table 2. Generally there is a better agreement between theory
and experiment than in the case of quadrupole excitations.
Other choices of the Skyrme forces do not improve the agreement obtained
with SIII.

An additional information  about the structure of the first $2^+, 3^-$
states can be extracted by looking at the ratio of the multipole
transition matrix elements $M_n/M_p$ that depend on the relative
contributions of the proton and neutron configurations. In the framework
of the collective model for isoscalar excitations this ratio is
equal to $M_n/M_p=N/Z$ and any deviation from this value can indicate
an isovector character of the state. The $M_n/M_p$ ratio can be determined
experimentally by using different external probes \cite{Ber83,Ken92,Jew99}.
Recently \cite{Khan00,Khan01}, QRPA calculations of
the $M_n/M_p$ ratios for the $2^+_1$ states in some O and Ar
isotopes have been done. The predicted results are in good
agreement with experimental data \cite {Khan01}. Our calculated values of
the $M_n/M_p$ ratios for the $2^+_1$ and $3^-_1$ states are shown in Tables
3 and 4, respectively.
Our results support the conclusions of Refs. \cite{Khan00,Khan01} about the
isovector character of the $2^+_1$ states in $^{18,20}$O and $^{38}$Ar.
As one can see from Table 4  our calculations predict that
the $M_n/M_p$ ratios for the $3^-_1$ states are rather close to $N/Z$,
thus indicating their isoscalar character.
This conclusion remains valid for the SGII and SkI4 parameter sets.

\section{Conclusion}

A finite rank separable approximation for the particle-hole RPA
calculations with Skyrme interactions that was proposed in
our previous work is extended to take into account the pairing correlations.
The QRPA equations are derived for this case. These equations are used
to study the evolution of quadrupole and octupole excitations in nuclei
away from stability.
It is shown that the suggested approach enables
one to reduce remarkably the dimensions of the matrices that
must be inverted to perform structure calculations in very large
configuration spaces.

As an illustration of the method we have used the finite rank p-h
interaction derived from the Skyrme force SIII to calculate the
energies and transition probabilities of the $2^+_1$ and $3^-_1$
states in some O, Ar, Sn and Pb isotopes.  The
values calculated within our approach are very close to those that were
calculated in QRPA with the full Skyrme interactions.
They are generally in a reasonable agreement with experimental data.
A further development will be to take into account
the coupling between the one- and two-phonon terms and such
investigations are in progress now.

\section{Acknowledgments}

A.P.S. and V.V.V. thank the hospitality of IPN-Orsay where the main part
of this work was done. This work is partly supported by IN2P3-JINR agreement
and by the Bulgarian Science Foundation (contract Ph-801).

\begin{appendix}

\section {}
For the sake of completeness we show how the finite rank form of the
residual forces (\ref{eq9})
can simplify the solution of the RPA equations (\ref{eq14}).
In the $4N$-dimensional space we introduce a vector
$\left(
\begin{array}{c}
\mathbf{{\cal D}}_M\left( \tau \right) \\
\mathbf{{\cal D}}_S\left( \tau \right)
\end{array}
\right)$
by its components:
\begin{equation}
\label{D}
\mathbf{{\cal D}}_\beta^k\left( \tau \right) =\left(
\begin{array}{c}
D_\beta^k\left( \tau \right) \\
D_\beta^k\left( -\tau \right)
\end{array}
\right),\beta=\left\{ M,S\right\}
\end{equation}
where
\[
D_M^{\lambda ik}\left( \tau \right) =\left. \sum_{jj^{^{\prime }}}\right.
^\tau f_{jj^{^{\prime }}}^{(\lambda k)}u_{jj^{^{\prime }}}^{\left( +\right)
}\left( X _{jj^{^{\prime }}}^{\lambda i}+ Y _{jj^{^{\prime
}}}^{\lambda i}\right),
\]
\[
D_S^{\lambda ik}\left( \tau \right) =\left. \sum_{jj^{^{\prime }}}\right.
^\tau f_{jj^{^{\prime }}}^{(\lambda \lambda k)}u_{jj^{^{\prime }}}^{\left(
-\right) }\left( X _{jj^{^{\prime }}}^{\lambda i}-Y _{jj^{^{\prime
}}}^{\lambda i}\right).
\]
The index k run over the $N$-dimensional space (k=1,2,...,N).
Solving the system of equations (\ref{eq14})  one can get the following expressions
for the phonon amplitudes:
\begin{equation}
X _{jj^{^{\prime }}}^{\lambda i}\left( \tau \right) =\frac 1{\left(
\varepsilon _{jj^{^{\prime }}}-\omega _{\lambda i}\right)
}\sum_{k=1}^{N}\frac 1{\sqrt{2{\cal Y}_\tau ^{\lambda ki}}}\left(
u_{jj^{^{\prime }}}^{\left( +\right) }f_{jj^{^{\prime }}}^{(\lambda
k)}+u_{jj^{^{\prime }}}^{\left( -\right) }f_{jj^{^{\prime }}}^{(\lambda
\lambda k)}z^{\lambda ik}\left( \tau \right) \right),
\label{AMPSI}
\end{equation}
\begin{equation}
\label{AMPHI}
Y _{jj^{^{\prime }}}^{\lambda i}\left( \tau \right) =\frac 1{\left(
\varepsilon _{jj^{^{\prime }}}+\omega _{\lambda i}\right)
}\sum_{k=1}^{N}\frac 1{\sqrt{2{\cal Y}_\tau ^{\lambda ki}}}\left(
u_{jj^{^{\prime }}}^{\left( +\right) }f_{jj^{^{\prime }}}^{(\lambda
k)}-u_{jj^{^{\prime }}}^{\left( -\right) }f_{jj^{^{\prime }}}^{(\lambda
\lambda k)}z^{\lambda ik}\left( \tau \right) \right),
\end{equation}
where
\[
{\cal Y}_\tau ^{\lambda ki}=\frac{2\left( 2\lambda +1\right) ^2}{\left(
D_M^{\lambda ik}\left( \tau \right) \left( \kappa _0^{\left( M,k\right)
}+\kappa _1^{\left( M,k\right) }\right) +D_M^{\lambda ik}\left( -\tau
\right) \left( \kappa _0^{\left( M,k\right) }-\kappa _1^{\left( M,k\right)
}\right) \right) ^2},
\]
\[
z^{\lambda ik}\left( \tau \right) =\frac{D_S^{\lambda ik}\left( \tau \right)
\left( \kappa _0^{\left( S,k\right) }+\kappa _1^{\left( S,k\right) }\right)
+D_S^{\lambda ik}\left( -\tau \right) \left( \kappa _0^{\left( S,k\right)
}-\kappa _1^{\left( S,k\right) }\right) }{D_M^{\lambda ik}\left( \tau
\right) \left( \kappa _0^{\left( M,k\right) }+\kappa _1^{\left( M,k\right)
}\right) +D_M^{\lambda ik}\left( -\tau \right) \left( \kappa _0^{\left(
M,k\right) }-\kappa _1^{\left( M,k\right) }\right) }.
\]
Using Eqs.(\ref{D}) and Eqs.(\ref{AMPSI}),(\ref{AMPHI})
the RPA equations (\ref{eq14})
can be reduced to the following system of equations:
\begin{equation}
\left(
\begin{array}{cc}
{\cal M}_{MM}\left( \tau \right) - 1 & {\cal M}_{MS}\left( \tau \right) \\
{\cal M}_{SM}\left( \tau \right) & {\cal M}_{SS}\left( \tau \right) -1
\end{array}
\right) \left(
\begin{array}{c}
\mathbf{{\cal D}}_M\left( \tau \right) \\
\mathbf{{\cal D}}_S\left( \tau \right)
\end{array}
\right) =0,
\end{equation}
where ${\cal M}$ is the $2N\times 2N$ matrix
\begin{equation}
{\cal M}_{\beta\beta^{\prime }}^{kk^{\prime }}\left( \tau \right) =\left(
\begin{array}{cc}
(\kappa _0^{\left( \beta^{\prime }, k^{\prime }\right) }+\kappa _1^{\left(
\beta^{\prime }, k^{\prime }\right) })T_{\beta\beta^{\prime }}^{kk^{\prime }}\left( \tau
\right)  & (\kappa _0^{\left( \beta^{\prime }, k^{\prime }\right) }-\kappa
_1^{\left( \beta^{\prime }, k^{\prime }\right) })T_{\beta\beta^{\prime }}^{kk^{\prime
}}\left( \tau \right)  \\
(\kappa _0^{\left( \beta^{\prime }, k^{\prime }\right) }-\kappa _1^{\left(
\beta^{\prime }, k^{\prime }\right) })T_{\beta\beta^{\prime }}^{kk^{\prime }}\left( -\tau
\right)  & (\kappa _0^{\left( \beta^{\prime }, k^{\prime }\right) }+\kappa
_1^{\left( \beta^{\prime }, k^{\prime }\right) })T_{\beta\beta^{\prime }}^{kk^{\prime
}}\left( -\tau \right)
\end{array}
\right),
\end{equation}

\[1 \leq k, k^{\prime } \leq N.\]

The matrix elements $T^{kk^{\prime }}$ have the following form:
\[
T_{MM}^{kk^{\prime }}\left( \tau \right) =\left. \sum_{jj^{^{\prime
}}}\right. ^\tau \frac{f_{jj^{^{\prime }}}^{(\lambda k)}f_{jj^{^{\prime
}}}^{(\lambda k^{\prime })}\left( u_{jj^{^{\prime }}}^{\left( +\right)
}\right) ^2\varepsilon _{jj^{^{\prime }}}}{\hat \lambda ^2\left( \varepsilon
_{jj^{^{\prime }}}^2-\omega _{\lambda i}^2\right) },
\]

\[
T_{MS}^{kk^{\prime }}\left( \tau \right) =\left. \sum_{jj^{^{\prime
}}}\right. ^\tau \frac{f_{jj^{^{\prime }}}^{(\lambda k)}f_{jj^{^{\prime
}}}^{(\lambda \lambda k^{\prime })}u_{jj^{^{\prime }}}^{\left( +\right)
}u_{jj^{^{\prime }}}^{\left( -\right) }\omega _{\lambda i}}{\hat \lambda
^2\left( \varepsilon _{jj^{^{\prime }}}^2-\omega _{\lambda i}^2\right) },
\]

\[
T_{SM}^{kk^{\prime }}\left( \tau \right) =T_{MS}^{k^{\prime }k}\left( \tau
\right),
\]

\[
T_{SS}^{kk^{\prime }}\left( \tau \right) =\left. \sum_{jj^{^{\prime
}}}\right. ^\tau \frac{f_{jj^{^{\prime }}}^{(\lambda \lambda
k)}f_{jj^{^{\prime }}}^{(\lambda \lambda k^{\prime })}\left( u_{jj^{^{\prime
}}}^{\left( -\right) }\right) ^2\varepsilon _{jj^{^{\prime }}}}{\hat \lambda
^2\left( \varepsilon _{jj^{^{\prime }}}^2-\omega _{\lambda i}^2\right) }.
\]

Thus, the RPA eigenvalues $\omega _{\lambda i}$ are the roots of the secular equation:
\begin{equation}
 \det\left(\begin{array}{cc}
{\cal M}_{MM}\left( \tau \right) - 1 & {\cal M}_{MS}\left( \tau \right) \\
{\cal M}_{SM}\left( \tau \right) & {\cal M}_{SS}\left( \tau \right) -1
\end{array}
\right) =0.
\end{equation}
The phonon amplitudes corresponding to the RPA eigenvalue $\omega _{\lambda i}$ are determined by
Eqs.(\ref{AMPSI}), (\ref{AMPHI}) taking into account the  normalization condition (\ref{eq12}).
\end{appendix}


\begin{table}[]
\caption[]{Energies and B(E2)-values for up-transitions to the first
$2^{+}$ states}
\begin{center}
\begin{tabular}{ccccc} 
Nucleus    & \multicolumn{2}{c} {Energy} & \multicolumn{2}{c} {B(E2$\uparrow$)}    \\
           & \multicolumn{2}{c} {(MeV)}  & \multicolumn{2}{c} {(e$^2$fm$^4$)}      \\
           & Exp. & Theory                & Exp.         & Theory                  \\
\hline
$^{18}$O   & 1.98 & 4.75   & 45$\pm$2          & 14                                \\
$^{20}$O   & 1.67 & 4.17   & 28$\pm$2          & 20                                \\
$^{36}$Ar  & 1.97 & 1.91   & 300$\pm$30        & 310                               \\
$^{38}$Ar  & 2.17 & 2.51   & 130$\pm$10        & 110                               \\
$^{40}$Ar  & 1.46 & 2.17   & 330$\pm$40        & 290                               \\
$^{112}$Sn & 1.26 & 1.49   & 2400$\pm$140      & 2600                              \\
$^{114}$Sn & 1.30 & 1.51   & 2400$\pm$500      & 2100                              \\
$^{206}$Pb & 0.80 & 0.96   & 1000$\pm$20       & 1700                              \\
$^{208}$Pb & 4.09 & 5.36   & 3000$\pm$300      & 2000                              \\
\end{tabular}
\end{center}
\end{table}

\begin{table} []
\caption[]{Energies and B(E3)-values for up-transitions to the first
 $3^{-}$ states}
\begin{center}
\begin{tabular}{ccccc}
Nucleus    & \multicolumn{2}{c} {Energy}  & \multicolumn{2}{c} {B(E3$\uparrow$)}   \\
           & \multicolumn{2}{c} {(MeV)}   & \multicolumn{2}{c} {(e$^2$fm$^6$)}     \\
           & Exp. & Theory                & Exp.               & Theory            \\
\hline
$^{18}$O   & 5.10 & 6.15                  &  1250 $ \pm$50     & 1200              \\
$^{20}$O   & 5.61 & 7.28                  &   530 $ \pm$180    & 1000              \\
$^{36}$Ar  & 4.18 & 4.26                  & 11100 $ \pm$1100   & 16000             \\
$^{38}$Ar  & 3.81 & 3.21                  & 10000 $ \pm$3000   & 15000             \\
$^{40}$Ar  & 3.68 & 4.85                  &  8700 $ \pm$1000   & 12000             \\
$^{112}$Sn & 2.36 & 2.73                  & 87000 $ \pm$12000  & 97000             \\
$^{114}$Sn & 2.28 & 2.31                  &100000 $ \pm$12000  & 97000             \\
$^{206}$Pb & 2.65 & 2.92                  &650000 $ \pm$40000  & 750000            \\
$^{208}$Pb & 2.62 & 2.66                  &611000 $ \pm$9000   & 860000            \\
\end{tabular}
\end{center}
\end{table}

\begin{table}[]
\caption[]{ $(M_n/M_p)/(N/Z)$ ratios for the first $2^{+}$ states}
\begin{center}
\begin{tabular}{cccccc} 
Nucleus    & $^{18}$O                & $^{20}$O       & $^{36}$Ar                & $^{38}$Ar & $^{40}$Ar         \\
\hline
Theory     &   2.1                   &  2.1               &  0.9                 & 0.5       & 0.9               \\
Exp.       &   0.88$\pm$0.19  $^a$   &  2.17$\pm$0.53 $^a$&  1.41$\pm$0.50 $^b$  &  --       & 0.68$\pm$0.21 $^b$\\
           &   1.62$\pm$0.02  $^c$   &  1.9$\pm$0.3 $^c$  & & & \\
\end{tabular}
\end{center}
{$^a$ Ref.\cite{Khan00},
 $^b$ Ref.\cite{Khan01},
 $^c$ Ref.\cite{Jew99}}
\end{table}

\begin{table}[]
\caption[]{ $(M_n/M_p)/(N/Z)$ ratios for the first $3^{-}$ states}
\begin{center}
\begin{tabular}{cccccc} 
Nucleus    & $^{18}$O  & $^{20}$O & $^{36}$Ar & $^{38}$Ar & $^{40}$Ar \\
\hline
Theory     &  0.8      &   0.8    & 0.9       & 1.0       &  0.9      \\
\end{tabular}
\end{center}
\end{table}

\begin{figure}
\caption {The multipole strength distributions in $^{36}$Ar}
\end{figure}
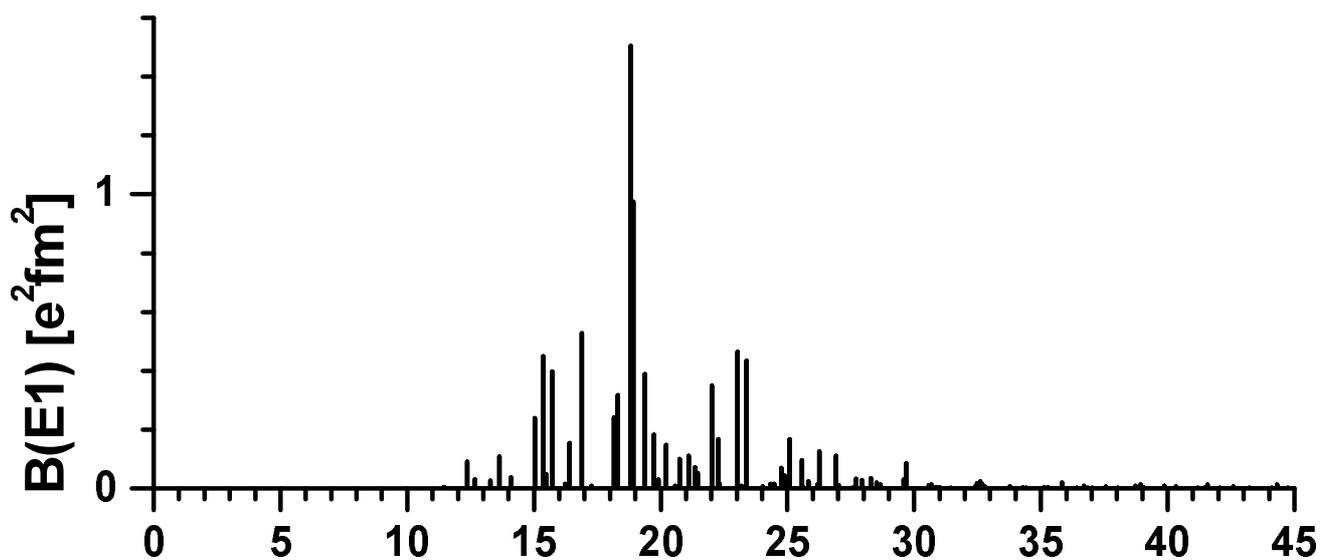
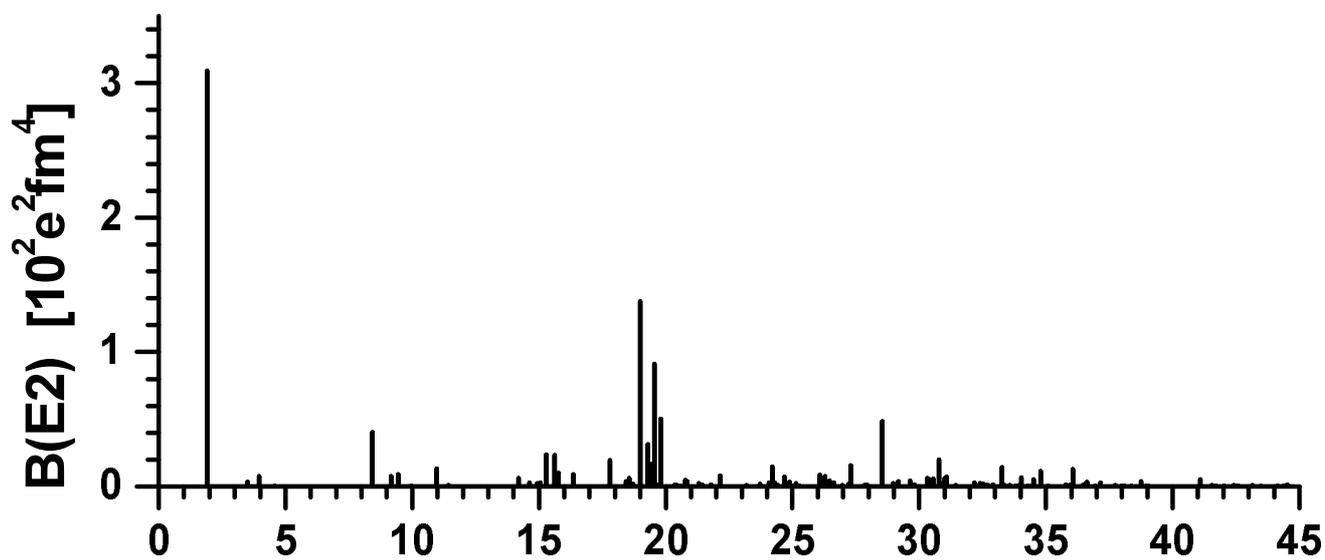
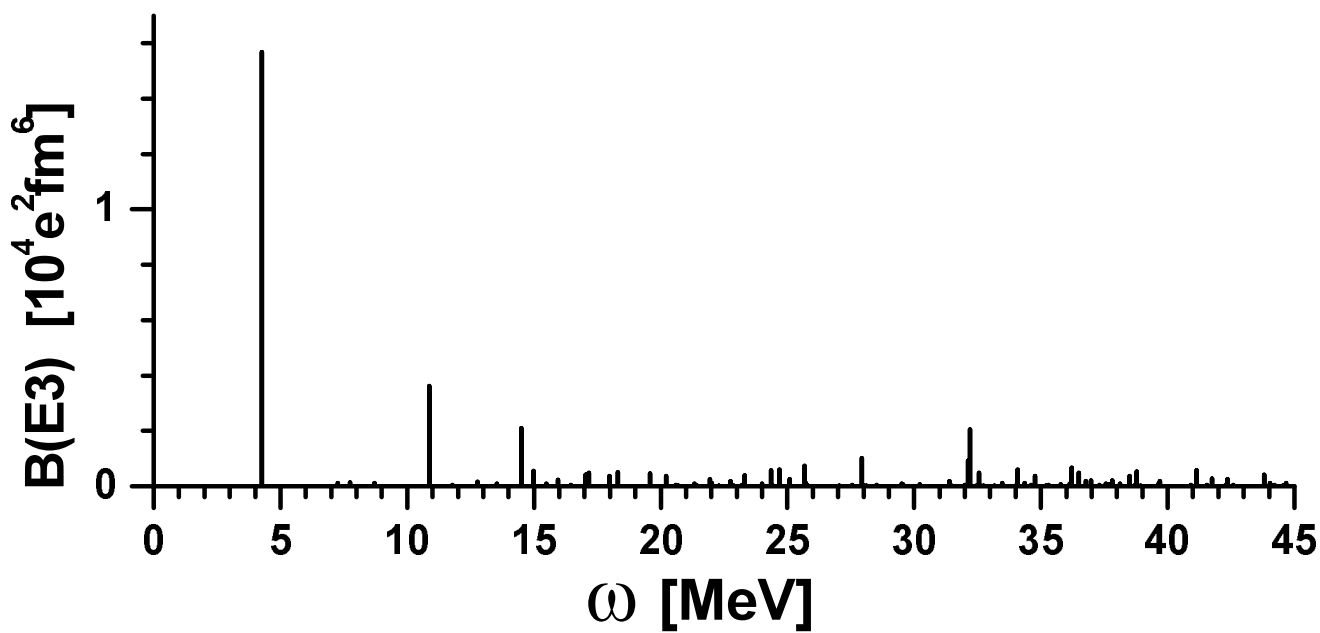
\end{document}